\documentclass[runningheads]{svmult}

\usepackage{makeidx}   % allows index generation
\usepackage{graphicx}  % standard LaTeX graphics tool
\usepackage{subeqnar}  % subnumbers individual equations
\usepackage{multicol}  % used for the two-column index
\usepackage{physprbb}  % modified textarea for proceedings,
\usepackage{psfrag}

\makeindex             % used for the subject index
\def\0{\varnothing}

\begin{document}
\title*{Ensemble Inequivalence in Mean-field Models
of Magnetism}
\toctitle{Ensemble Inequivalence in Mean-field
Models of Magnetism}
% allows explicit linebreak for the table of content
%
%
\titlerunning{Ensemble Inequivalence}
% allows abbreviation of title, if the full title is too long
% to fit in the running head
%
\author{Julien Barr{\'e} \inst{1,3}
\and David Mukamel \inst{2}
\and Stefano Ruffo \inst{1,3}}
\authorrunning{Julien Barr{\'e} et al.}
% if there are more than two authors,
% please abbreviate author list for running head
%
%
\institute{ Ecole Normale Sup{\'e}rieure de Lyon, Laboratoire
de Physique, 46 All{\'e}e d' Italie, 69364 Lyon Cedex 07, France
\and Department of Physics of Complex Systems, The Weizmann
    Institute of Science, Rehovot 76100, Israel
\and Dipartimento di Energetica ``Sergio Stecco'', Universit{\`a}
    di Firenze, via s. Marta 3 50139 Firenze, Italy, INFM and INFN.}

\maketitle              % typesets the title of the contribution

\begin{abstract}
Mean-field models, while they can be cast into an {\it extensive}
thermodynamic formalism, are inherently {\it non additive}. This
is the basic feature which leads to {\it ensemble inequivalence}
in these models. In this paper we study the global phase diagram
of the infinite range Blume-Emery-Griffiths model both in the {\it
canonical} and in the {\it microcanonical} ensembles. The
microcanonical solution is obtained both by direct state counting
and by the application of large deviation theory. The canonical
phase diagram has first order and continuous transition lines
separated by a tricritical point. We find that below the
tricritical point, when the canonical transition is first order,
the phase diagrams of the two ensembles disagree. In this region
the microcanonical ensemble exhibits energy ranges with negative
specific heat and temperature jumps at transition energies. These
two features are discussed in a general context and the
appropriate Maxwell constructions are introduced. Some preliminary
extensions of these results to weakly decaying nonintegrable
interactions are presented.
\end{abstract}

\section{Introduction}
\label{Intro} Thermodynamics of systems with long range
interactions is quite distinct from that of systems where the
interactions are short ranged. For pairwise interactions, long
range potentials decay at large distances as $V(r)\sim
1/r^{\alpha}$ with $0 \leq \alpha \leq d$ in $d$ dimensions. Such
systems are {\it not extensive} \index{extensivity}, although models
used to describe
them can be made extensive by an appropriate volume dependent
rescaling (the Kac prescription)~\cite{Kac} \index{Kac prescription}.
However, even with
such rescaling, these models remain inherently
 {\it non-additive}\index{additivity}.
Namely, when divided into two or more macroscopic subsystems, the
total energy (or other extensive quantity) of the system, is not
necessarily equal to the sum of the energies of the subsystems.
Since additivity is an essential ingredient in the derivation of
thermodynamics and statistical mechanics \index{statistical mechanics}
of systems with short
range interactions, its violation may lead to interesting and
unusual effects.

\begin{figure}[]
\begin{center}
\includegraphics[width=.5\textwidth]{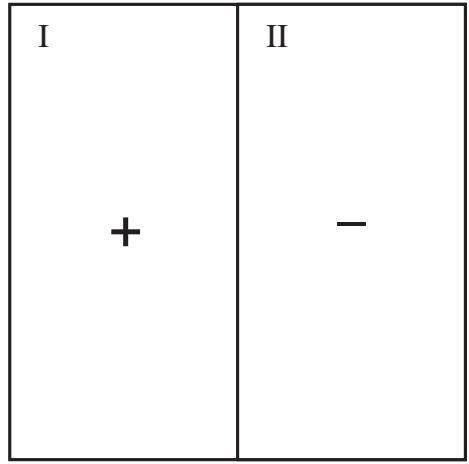}
\end{center}
\caption[]{A two phase configuration of the infinite range
interaction Ising model for which the energy is non-additive}
\label{plusminus}
\end{figure}

To illustrate the lack of additivity in these systems consider an
extremely simple example of an Ising model with infinite range,
mean-field like, interactions, corresponding to 
$\alpha=0$\index{mean field!models}. The
Hamiltonian takes the form
\begin{equation}
H= -\frac{J}{2N}\left(\sum_{i=1}^N \sigma_i\right)^2
\label{Ising}
\end{equation}
where $\sigma_i =\pm 1$ is the spin $1/2$ variable on site $i$ and
$N$ is the number of spins in the system. This Kac \index{Kac prescription}
type pre-factor
is taken to ensure {\it extensivity}\index{extensivity}. Let us consider for
simplicity a system with vanishing magnetization, $M=\sum_{i=1}^N
\sigma_i=0$. It is clear that the energy, $E=H$, of the system
satisfies $E=0$. Now let us divide the system into two subsystems,
$I$ and $II$, each composed of $N/2$ sites, where all spins in
subsystem $I$ are up while those in subsystem $II$ are down (see
Fig.~\ref{plusminus}). The energies of the two subsystems, $E_I$ and $E_{II}$
satisfy $E_I=E_{II}=-JN/4$. Since the sum of the two energies
$E_I+E_{II}$ is not equal to the total energy $E$ the system is
clearly not additive.

\begin{figure}[]
\begin{center}
\includegraphics[width=.5\textwidth]{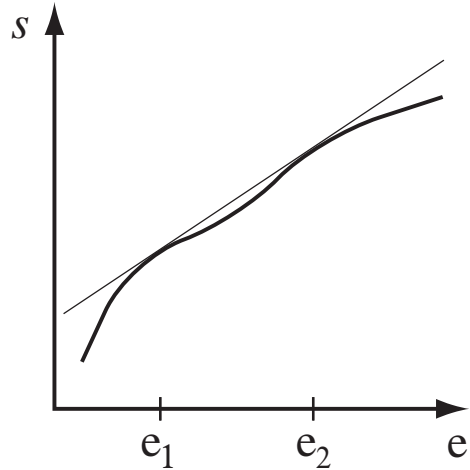}
\end{center}
\caption[]{A non-concave entropy curve, which for additive systems
is made concave by considering two phase coexistence
configurations in the region where the entropy of the homogeneous
system is non-concave. This procedure is not applicable for
non-additive systems like those with long range interactions.}
\label{convex}
\end{figure}

The lack of additivity \index{additivity} may result in many
unusual properties of systems with long range interactions, like
inequivalence of various ensembles\index{ensemble inequivalence},
negative specific heat \index{negative!specific heat} in the
microcanonical ensemble \index{microcanonical!ensemble} and
possible temperature discontinuity \index{temperature
discontinuity} at first order transitions. For example, the usual
argument for the non-negativity of the specific heat, or
equivalently, for the concavity \index{convexity} of the
entropy-energy curve $s(\epsilon)$, makes use of the fact that
systems with short range interactions are additive. Here $s=S/N$
and $e=E/N$ are the entropy and energy per particle, respectively.
An entropy curve which is not concave in the energy interval
$e_1<e<e_2$ (see Fig.~\ref{convex}) is unstable for systems with
short range interactions since entropy may be gained by phase
separating the system into two subsystems with energy densities
$e_1$ and $e_2$ while keeping the total energy fixed. In
particular the average energy and entropy densities in the
coexistence region is given by the weighted average of the
corresponding densities of the two coexisting systems, and thus
the correct entropy curve in that region is given by the common
tangent line (see Fig.~\ref{convex}), resulting in an overall
concave curve. On the other hand for systems with long range
interactions, additivity does not hold, and thus it is not always
the case that entropy may be gained by phase separation. In such
cases the non-concave entropy curve may in fact be the correct
entropy of the system, resulting, for example, in negative
specific heat\index{negative!specific heat}.

Self gravitating \index{gravitation} particles are perhaps the most
extensively studied systems with long range forces
(corresponding to $\alpha =1$ in $d=3$ dimensions).
The fact that the entropy of these systems
is not necessarily a concave function of the energy has first been
pointed out by Antonov~\cite{Antonov}. The thermodynamical
consequences of this observation have been elaborated using simple
models by Lynden-Bell~\cite{Lynden}, Thirring~\cite{Thirring} and
others (see also the papers by Chavanis~\cite{Chavanis} and
Padhmanaban~\cite{Padhmanaban} in this book for a review).

In fact, non-additivity \index{additivity} is not limited to systems with
long range interactions. Finite systems with short range
interactions, such as atomic clusters, are non-additive as long as
they are small enough so that surface energy may not be neglected
as compared with the bulk energy~\cite{Gross}. Negative specific heat in
\index{negative!specific heat} clusters of atoms has been discussed by
R. M. Lynden-Bell~\cite{RMBell} and it has been recently observed
experimentally, as discussed in this book in Refs.~\cite{Gross,Chomaz}.

In the present paper we consider a simple spin 1 mean-field model
\index{mean field!models} ($\alpha = 0$) and study its phase
diagram both within the canonical and the microcanonical
ensembles. This model, known as the Blume-Emery-Griffiths (BEG)
model~\cite{Blume}, \index{Blume Emery Griffiths (BEG) model} may
serve as a useful tool for studying thermodynamic effects which
characterize systems with long range interactions. The canonical
phase diagram \index{canonical!ensemble}of this model has been
studied in the past, and it exhibits a line of phase transitions
separating a magnetically ordered phase from a paramagnetic one.
The line is composed of a first order transition and a second
order transition segments separated by a tricritical
 point\index{tricritical point}. The microcanonical phase diagram
\index{microcanonical!ensemble} on the other hand exhibits some
rather distinct features~\cite{BMR}. While the second order
segment of the canonical phase diagram is present also in the
microcanonical ensemble, the nature of the first order segment,
its location and the location of the tricritical point are rather
different. In this ensemble, a region with negative specific heat
is found, as well as a first order transition at which the
temperatures of the two coexisting phases are different.

The paper is organized as follows: in Sec.~\ref{Section2} the BEG
model is introduced and its canonical phase diagram is studied.
The microcanonical phase diagram and the relationship between the
two ensembles are studied in Sec.~\ref{Section3}. Some features of
the Maxwell constructions leading to temperature jumps are
presented in Sec.~\ref{Section4}. An alternative more general
derivation of the expression for microcanonical entropy is
presented in Sec.~\ref{Section5}. The results presented in this
paper are generalized to models with slowly decaying interactions
(with $0 < \alpha <d$) in Sec.~\ref{Section6}. Finally, concluding
remarks are given in Sec.~\ref{Section7}

\section{The Blume-Emery-Griffiths model in the canonical ensemble}
\label{Section2}
\begin{figure}
\centering

\psfrag{-1}[Bl][Bl][0.7][0]{-1}
\psfrag{1}[Bl][Bl][0.7][0]{1}
\psfrag{0.5}[Bl][Bl][0.7][0]{}
\psfrag{-0.5}[Bl][Bl][0.7][0]{}
\psfrag{0}[Bl][Bl][0.7][0]{0}
\psfrag{1.3}[Bl][Bl][0.7][0]{1.3}
\psfrag{1.4}[Bl][Bl][0.7][0]{1.4}
\psfrag{1.5}[Bl][Bl][0.7][0]{1.5}
\psfrag{1.6}[Bl][Bl][0.7][0]{1.6}
\psfrag{1.7}[Bl][Bl][0.7][0]{1.7}
\psfrag{-1.1}[Bl][Bl][0.7][0]{-1.1}
\psfrag{-1.05}{}
\normalsize
\psfrag{betaf}{$\mathbf{\beta \tilde{f}}$}
\psfrag{x}{\bf{x}}
\psfrag{betaJ}{$\mathbf{\beta J}$}
\includegraphics[width=.7\textwidth]{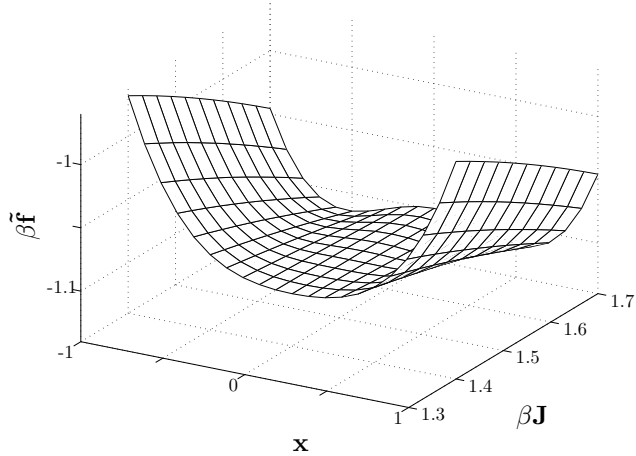}
\caption[]{$\beta \tilde{f}(\beta,x)$ as a function of $\beta J\in[1.4,1.7]$
and $x\in[-1,1]$ for
$\Delta=0$. The second order transition is at $\beta J=3/2$ where
the minimum at $x=0$ becomes a maximum and two side
minima develop}
\label{secondorder}
\end{figure}

\begin{figure}
\centering
\psfrag{betaf}{$\mathbf{\beta \tilde{f}}$}
\psfrag{x}{\bf{x}}
\psfrag{betaJ}{$\mathbf{\beta J}$}
\psfrag{-1}[Bl][Bl][0.7][0]{-1}
\psfrag{1}[Bl][Bl][0.7][0]{1}
\psfrag{0.5}[Bl][Bl][0.7][0]{}
\psfrag{-0.5}[Bl][Bl][0.7][0]{}
\psfrag{0}[Bl][Bl][0.7][0]{0}
\psfrag{3.8}[Bl][Bl][0.7][0]{3.8}
\psfrag{4}[Bl][Bl][0.7][0]{4}
\psfrag{4.2}[Bl][Bl][0.7][0]{4.2}
\psfrag{4.4}[Bl][Bl][0.7][0]{4.4}
\psfrag{-0.2}[Bl][Bl][0.7][0]{-0.2}
\psfrag{-0.3}[Bl][Bl][0.7][0]{-0.3}
\psfrag{-0.18}{}
\psfrag{-0.22}{}
\psfrag{-0.24}{}
\psfrag{-0.26}{}
\psfrag{-0.28}{}
\psfrag{3.7}{}
\psfrag{3.9}{}
\psfrag{4.1}{}
\psfrag{4.3}{}
\psfrag{4.5}{}
\psfrag{1.7}[Bl][Bl][0.7][0]{1.7}
\psfrag{-1.1}[Bl][Bl][0.7][0]{-1.1}
\psfrag{-1.05}{}

\includegraphics[width=.7\textwidth]{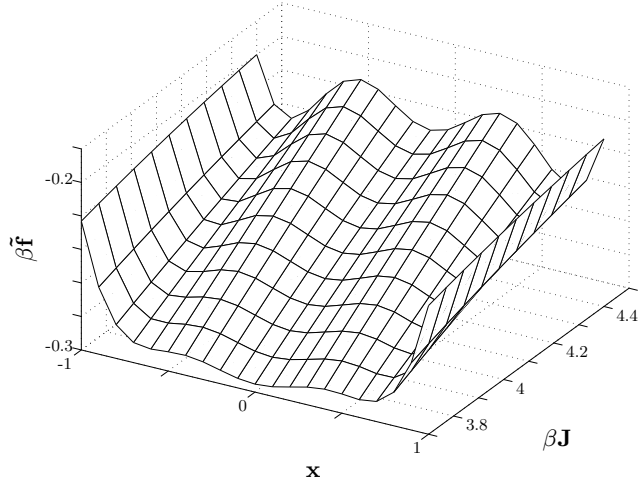}
\caption[]{$\beta \tilde{f}(\beta,x)$ as a function of $\beta J\in[3.7,4.4]$
and $x\in[-1,1]$ for
$\Delta=0.476190$. The first order transition is at $\beta J=4$.
Two side minima to $x=0$ first develop and then reach the
same hight at the transition point.}
\label{first}
\end{figure}

\index{canonical!ensemble}
We consider a spin model with infinite range, mean-field like,
interactions whose phase diagram can be analyzed analytically both
within the canonical and the microcanonical ensembles. This study
enables one to compare the two resulting diagrams and get a better
understanding of the effect of the non-additivity on the
thermodynamic behavior of the system. The model we consider is a
simple version of the Blume-Emery-Griffiths (BEG) model~\cite{Blume}, known as
the Blume-Capel model, with infinite range interactions. The model
is defined on a lattice, where each lattice point $i$ is occupied
by a spin-1 variable $S_i=0,\pm 1$. The Hamiltonian is given by
\begin{equation}
H=\Delta\sum_{i=1}^N S_i^2 -\frac{J}{2N}\left(\sum_{i=1}^N S_i\right)^2
\label{BEG}
\end{equation}
where $J>0$ is a ferromagnetic coupling constant and $\Delta$
controls the energy difference between the magnetic $(S_i=\pm1)$
and the non-magnetic $(S_i=0)$ states. The canonical phase diagram
of this model \index{canonical!ensemble} in the $(T,\Delta)$ plane has been studied in the
past~\cite{Blume}.
The free energy
\begin{equation}
f(\beta)=-\frac{1}{\beta}\lim_{N\to\infty}\frac{\ln Z}{N},
\label{free-energy}
\end{equation}
where $Z$ is the partition function
\begin{equation}
Z=\sum_{[S_i]}\exp{-\beta H},
\label{partition}
\end{equation}
with $\beta=1/{k_B T}$ and $k_B$ the Boltzmann constant,
can be exactly derived in the $N\to\infty$ limit.
One uses the Hubbard-Stratonovich transformation, which
in this case amounts to the simple Gaussian identity
\begin{equation}
\exp(ba^2)=\sqrt{\frac{b}{\pi}}\int_{-\infty}^{\infty} dx \exp(-bx^2+2abx),
\end{equation}
with $a=\sum_iS_i/N$ and $b=\beta JN/2$. One then easily gets
\begin{equation}
Z=\sqrt{\frac{\beta JN}{2 \pi}}\int_{-\infty}^{\infty} dx 
\exp(-N\beta \tilde{f}(\beta,x))
\label{part}
\end{equation}
where
\begin{equation}
\beta \tilde{f}(\beta,x)={1 \over 2}\beta Jx^2
-\ln[1+e^{-\beta \Delta}(e^{\beta Jx}+e^{-\beta Jx})].
\label{free}
\end{equation}
Using Laplace's method to perform the integral
in formula (\ref{part}) in the $N\to\infty$ limit, one finally
gets
\begin{equation}
f(\beta)=\min_x \tilde{f}(\beta,x)
\end{equation}
At $T=0$ the model has a ferromagnetic phase for $2\Delta /J<1$
and a non-magnetic phase otherwise separated by a first order
phase transition. Indeed, the paramagnetic zero temperature state
$S_i=0,\forall i$ is degenerate with the ferromagnetic state
$S_i=1,\forall i$ (or $S_i=-1,\forall i$) at  $2\Delta /J=1$ (this
latter state being the ground state for $2\Delta /J<1$), a typical
scenario for first order phase transitions. At $\Delta=-\infty$
the model reduces to the mean-field Ising model and hence it has a
second order phase transition at ${k_B}T/J=1$. The $(T, \Delta)$
phase diagram displays a transition line separating the low
temperature ferromagnetic phase from the high temperature
paramagnetic phase. The transition line is found to be first order
at high $\Delta$ values, while it is second order at low $\Delta$.
The second order line is found by locating the instability of the
paramagnetic phase $x=0$, which means finding the condition for
the $x^2$ coefficient in formula (\ref{free}) to be zero (note
that $\tilde{f}(\beta,x)$ is even in $x$). One gets for the second order
term
\begin{equation}
\beta J = {1 \over 2} e^{\beta \Delta} +1~.
\label{CriticaLine}
\end{equation}
The two segments of the transition line (high and low $\Delta$)
are separated by a tricritical point located at $\Delta/J=\ln(4)/3
\simeq0.4621$, $\beta J=3$.
This is obtained by requiring that both the $x^2$ coefficient of
$\tilde{f}(\beta,x)$ in (\ref{free}) and the $x^4$ coefficient
\begin{equation}
\frac{1}{3}-\frac{2 \exp(-\beta\Delta)}{(1+2\exp(-\beta\Delta)}
\end{equation}
vanish. The first order segment of the transition line is obtained
numerically by equating the free energies of the ferromagnetic and
the paramagnetic states. The behavior of the function $\beta
\tilde{f}(\beta,x)$ as $\beta$ varies is shown in Fig. \ref{secondorder} at
a second order phase transition ($\Delta=0$) and in
Fig. \ref{first} at a first order phase transition
($\Delta=0.476190$).

\section{Microcanonical solution of the Blume-Emery-Griffiths model}
\label{Section3}
\begin{figure}[]
\begin{center}
\includegraphics[width=.5\textwidth]{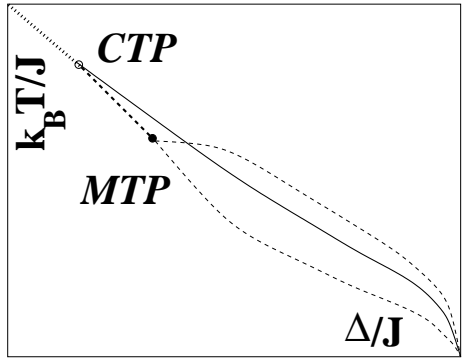}
\end{center}
\caption[]{A schematic representation of the phase diagram, where
we expand the region around the canonical (CTP) and the
microcanonical (MTP) tricritical points. The second order line,
common to both ensembles, is dotted, the first order canonical
transition line is solid and the microcanonical transition lines
are dashed (with the bold dashed line representing a continuous
transition)} \label{schematic}
\end{figure}

\begin{figure}[]
\begin{center}
\includegraphics[width=.5\textwidth,angle=-90]{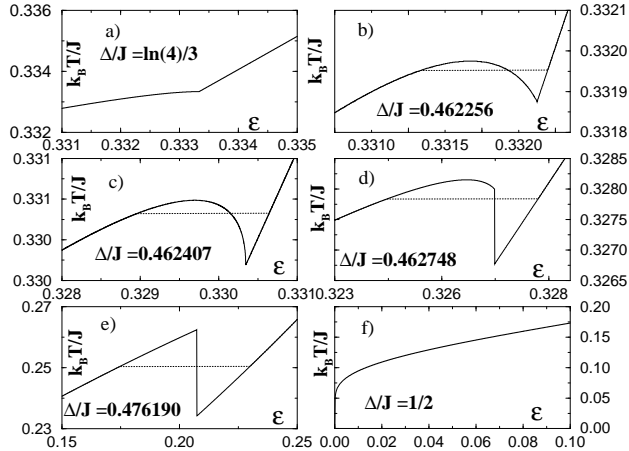}
\end{center}
\caption[]{The temperature-energy relation in the microcanonical
ensemble for different values of $\Delta$. The dotted horizontal
line in some of the plots is the Maxwell construction
\index{Maxwell!construction} in the canonical ensemble and
identifies the canonical first order transition temperature at the
point where two minima of the free energy coexist. The vertical
line in panels d) and e) marks the presence of a temperature jump:
for this energy two entropy maxima of the same height coexist}
\label{tvse}
\end{figure}

\index{microcanonical!ensemble}
The derivation of the phase diagram of the BEG
model (\ref{BEG}) in the {\it microcanonical ensemble} reduces to
a simple counting problem, since all spins interact with equal
strength, independently of their mutual distance (there is no
space in the model). A given microscopic configuration is
characterized by the numbers $N_+, N_-, N_0$ of up, down and zero
spins, with $N_+ + N_- + N_0 = N$. The energy $E$ of this
configuration is only a function of $N_+, N_-$ and $N_0$ and is
given by
\begin{equation}
\label{Energy}
E=\Delta Q - {J \over {2N}}M^2~,
\end{equation}
where $Q=\sum_{i=1}^N S_i^2=N_+ + N_-$ (the quadrupole moment) and
$M=\sum_{i=1}^N S_i=N_+ - N_-$ (the magnetization) are the two order
parameters. The number of microscopic configurations
$\Omega$ compatible with the macroscopic occupation numbers
$N_+, N_-$ and $N_0$ is
\begin{equation}
\label{Omega}
\Omega = {{N!} \over {{N_+!}{N_-!}{N_0!}}}~.
\end{equation}
Using Stirling's approximation in the large $N$ limit,
the entropy, $S=k_B \ln \Omega$, is given by
\begin{eqnarray}
\label{Entropy1}
S&=&-k_B N [(1-q)\ln (1-q) + {1 \over 2}(q+m) \ln (q+m) \nonumber \\
&+&{1 \over 2}(q-m) \ln (q-m) - q \ln 2]~,
\end{eqnarray}
where $q=Q/N$ and $m = M/N$ are the quadrupole moment and the
magnetization per site, respectively. Let $\epsilon = E/{\Delta
N}$ be the dimensionless energy per site, normalized by $\Delta$.
Equation (\ref{Energy}) may be written as
\begin{equation}
q = \epsilon +Km^2~,
\label{qofm}
\end{equation}
where $K=J/{2\Delta}$. Using this relation, the entropy per site
$s=S/(k_B N)$ can be expressed in terms of $m$ and $\epsilon$. At
fixed $\epsilon$, the value of $m$ which maximizes the entropy
corresponds to the equilibrium magnetization. The corresponding
equilibrium entropy $s(\epsilon)=\max_m s(\epsilon,m)$ contains
all the information about the thermodynamics of the system in the
microcanonical ensemble. For instance, temperature can be obtained
from the relation
\begin{equation}
\frac{\Delta}{k_B T}= \frac{\partial s}{\partial \epsilon}~.
\label{temp}
\end{equation}
As usual in systems where the energy per particle is bounded from
above, the model has both a positive and a negative temperature
region: entropy is a one humped function of the energy. The
interesting features take place in the positive temperature range.
In order to locate the continuous transition line, one develops
$s(\epsilon,m)$ in powers of $m$, in analogy with what has been
done above for the canonical free energy
\begin{equation}
\label{EntropyExpansion}
s=s_0 +Am^2 + Bm^4 + O(m^6)~,
\end{equation}
where
\begin{equation}
\label{s0}
s_0=s(\epsilon,m=0)=-(1-\epsilon)\ln(1-\epsilon) - \epsilon \ln \epsilon +\epsilon \ln 2~,
\end{equation}
and
\begin{eqnarray}
\label{AB}
A&=&-K \ln {\epsilon \over {2(1-\epsilon)}} -{1 \over {2 \epsilon}}~,
\nonumber \\
B&=&-{K^2 \over {2 \epsilon(1-\epsilon)}}+{K \over {2
\epsilon^2}}-{1\over {12 \epsilon^3}}~.
\end{eqnarray}
In the paramagnetic phase both $A$ and $B$ are negative, and the
entropy is maximized by $m=0$. The continuous transition to the
ferromagnetic phase takes place at $A=0$ for $B<0$. In order to
obtain the critical line in the $(T,\Delta)$ plane we first observe that
temperature is calculable on the critical line ($m=0$)
using (\ref{temp}) and (\ref{s0})
\begin{equation}
\label{Temperature}
\frac{\Delta}{k_B T} = \ln {{2(1- \epsilon)} \over \epsilon}~.
\end{equation}
Requiring now that $A=0$, one gets the following expression
for the critical line
\begin{equation}
\label{MicroCritical}
2 \bar\beta K ={1 \over 2} e^{\bar\beta} +1~,
\end{equation}
where $\bar\beta \equiv \beta \Delta$.
Equivalently, this expression may be written as
$\bar\beta K = 1 /{2\epsilon}$. The microcanonical critical line
thus coincides with the critical line (\ref{CriticaLine}) obtained
for the canonical ensemble.
The tricritical point of the microcanonical ensemble is obtained
at $A=B=0$. Combining these equations with Eq.~(\ref{Temperature})
one finds that at the tricritical point $\bar\beta$ satisfies
\begin{equation}
\label{MicroTricritical}
{1 \over {8 {\bar\beta}^2}}{{e^{\bar\beta} +2}\over {e^{\bar\beta}}} -{1\over {4
{\bar\beta}}} + {1 \over 12} =0~.
\end{equation}
Equations (\ref{MicroCritical}, \ref{MicroTricritical}) yield a
tricritical point at $K \simeq 1.0813$, $\bar\beta \simeq 1.3998$.
This has to be compared with the canonical tricritical point
located at $K=3/ \ln (16) \simeq 1.0820$, $\bar\beta = \ln (4)
\simeq1.3995$. The two points, although very close to each other,
do not coincide. The microcanonical critical line extends beyond
the canonical one. In the region between the two tricritical
points, the canonical ensemble yields a first order transition at
a higher temperature, while in the microcanonical ensemble the
transition is still continuous. It is in this region, as discussed
below, that negative heat capacity \index{negative!specific heat}
appears. A schematic phase
diagram near the canonical tricritical point (CTP) and the
microcanonical one (MTP) is given in Fig.~\ref{schematic}. Beyond
the microcanonical tricritical point the temperature has a jump at
the transition energy in the microcanonical ensemble. The two
lines emerging on the right side from the MTP correspond to the
two limiting temperatures which are reached when approaching the
transition energy from below and from above (see
Fig.~\ref{tvse}d). The two microcanonical temperature lines and
the canonical first-order transition line all merge on the $T=0$
line at $2\Delta/J=1$.

To get a better undertanding of the microcanonical phase diagram
and also to compare our results with those obtained for
self-gravitating and for finite systems we consider the
temperature-energy relation $T(\epsilon)$. This curve has two
branches: a high energy branch (\ref{Temperature}) corresponding
to $m=0$, and a low energy branch obtained from (\ref{temp}) using
the spontaneous magnetization $m_s(\epsilon)$. At the intersection
point of the two branches the two entropies become equal. Their
first derivatives at the crossing point can be different,
resulting in a jump in the temperature, i.e. {\it a microcanonical
first order transition}. When the transition is continuous in the
microcanonical ensemble, i.e. the first derivative of the entropy
branches at the crossing point are equal, our model always
displays, at variance with what happens for gravitational systems,
a discontinuity in the second derivative of the entropy. This is
due to the fact that here we have a true symmetry breaking
transition. Fig.~\ref{tvse} \index{caloric curve} displays the
$T(\epsilon)$ curve for increasing values of $\Delta$. For $\Delta
/J =\ln(4)/3)$, corresponding to the canonical tricritical point,
the lower branch of the curve has a zero slope at the intersection
point (Fig.~\ref{tvse}a). Thus, the specific heat of the ordered
phase diverges at this point. This effect signals the canonical
tricritical point as it appears in the microcanonical ensemble.
Increasing $\Delta$ to the region between the two tricritical
points a {\it negative specific heat} \index{negative!specific heat} 
in the microcanonical ensemble first arises ($\partial
T/\partial \epsilon <0$), see Fig.~\ref{tvse}b. At the
microcanonical tricritical point $\Delta$ the derivative $\partial
T/\partial \epsilon$ of the lower branch diverges at the
transition point, yielding a vanishing specific heat
(Fig.~\ref{tvse}c). For larger values of $\Delta$ a jump in the
temperature appears at the transition energy (Fig.~\ref{tvse}d).
The lower temperature corresponds to the $m=0$ solution
(\ref{Temperature}) and the upper one is given by $\exp(\bar
\beta)= 2(1-q^*)/\sqrt{(q^*)^2-(m^*)^2}$, where $m^*,q^*$ are the
values of the order parameters of the ferromagnetic state at the
transition energy. The negative specific heat branch disappears at
larger values of $\Delta$, leaving just a temperature jump (see
Fig.~\ref{tvse}e). In the $\Delta/J \to 1/2$ limit the low
temperature branch, corresponding to $q=m=1$ in the limit, shrinks
to zero and the $m=0$ branch (\ref{Temperature}) occupies the full
energy range (Fig.~\ref{tvse}f).

We now present some general considerations concerning the
intricate relation between canonical and microcanonical ensembles.
The following discussion is quite general, and not specific to the
BEG model. Let us first calculate the canonical partition sum
using the density of states $\Omega=e^{Ns(e,m)}$, where $e=E/N$ is
the energy density and $m$ the order parameter (for the BEG model
$e=\epsilon\Delta$, $m$ is the magnetization and $q$ is obtained from
$e$ and $m$ by relation (\ref{qofm})).
\begin{eqnarray}
Z &=& \sum_{[S_i]}\exp\left(-\beta H({S_i})\right)\nonumber \\
           &=& \sum_{[e,m]}e^{Ns(e,m)}e^{-N\beta e}~,
\label{zofomega}
\end{eqnarray}
where we have replaced the sum over the configurations by the sum
over the values of the energy per spin and the order parameter,
which for finite $N$ take discrete values. In the thermodynamic
limit $N\to\infty$, the free energy in the canonical ensemble is
thus given by
\begin{equation}
\label{canoproblem}
f(\beta)=\min_{e,m}\left(e -\frac{s(e,m)}{\beta}\right)~.
\end{equation}
The point where the minimum is reached yields the mean energy and
magnetization per spin in the canonical ensemble.
As explained above, the microcanonical entropy is
\begin{equation}
\label{microproblem}
s(e)=\max_{m}s(e,m)
\end{equation}
and the magnetization in the microcanonical ensemble is the point
where the maximum is reached.\\
All questions are now reduced to the solution of the two variational problems
(\ref{canoproblem}) and (\ref{microproblem}).
The first order conditions are the same for the two problems
\begin{eqnarray}
\label{variational}
\frac{\partial s}{\partial m}&=&0 \nonumber\\
\frac{\partial s}{\partial e}&=&\beta,
\end{eqnarray}
where the second condition is indeed given in the microcanonical case
by the definition of temperature.
We will denote by $e^*(\beta),m^*(\beta)$ the solution of the variational problem
(\ref{variational}). Using (\ref{variational}), it is straightforward
to verify that
\begin{equation}
\frac{d(\beta f)}{d\beta}=e^*(\beta),
\end{equation}
meaning that the canonical mean energy coincides with the
minimizing energy. Hence, the extrema in (\ref{canoproblem}) and
in (\ref{microproblem}) are the same. However this does not imply
ensemble
equivalence: we have to study the stability of these extrema.\\

In order to discuss the stability of the canonical solution one
has to determine the sign of the eigenvalues of the Hessian of the
function to be minimized in (\ref{canoproblem}). The Hessian is
\begin{equation}
{\cal H}=\frac{-1}{\beta}
\left( \begin{array}{cc}
s_{mm} & s_{me} \\
s_{em} & s_{ee}
\end{array} \right)
\end{equation}
where, for example, $s_{mm}$ is the second derivative of $s$ with
respect to $m$. The extremum is a minimum if and only if the
determinant and the trace of the Hessian are positive
\begin{eqnarray}
-s_{ee}-s_{mm}&>&0 \\
s_{ee}s_{mm}-s_{me}^2&>&0,
\label{condstab}
\end{eqnarray}
which implies that $s_{ee}$ and $s_{mm}$ must be negative, and
moreover $s_{ee}<-s_{me}^2/|s_{mm}|$. This has strong
implications on the canonical specific heat, which must be
positive. Let us prove it using the variational approach, instead
of the usual Thirring argument~\cite{Thirring}, that uses the
expression of the canonical partition sum. Indeed, taking the
derivatives of Eqs.~(\ref{variational}) with respect to $\beta$,
after having substituted into them $e^*(\beta),m^*(\beta)$, one
gets
\begin{eqnarray}
s_{ee}\frac{de^*}{d\beta}+s_{em}\frac{dm^*}{d\beta}=1\\
s_{me}\frac{de^*}{d\beta}+s_{mm}\frac{dm^*}{d\beta}=0,
\label{second}
\end{eqnarray}
where it is understood that all second derivatives are computed at
$e^*(\beta),m^*(\beta)$. Recalling now that the specific heat per
particle at constant volume is,
\begin{equation}
\frac{C_V}{Nk_B}=\frac{de^*}{dT}=-\beta^2\frac{de^*}{d\beta}
\end{equation}
one gets
\begin{equation}
\frac{C_{V}}{Nk_B}=\beta^2\frac{s_{mm}}{s_{em}^2-s_{ee}s_{mm}},
\label{cv}
\end{equation}
which is always positive if the stability conditions
(\ref{condstab}) are satisfied. However, since the stability
condition in the microcanonical ensemble only requires that
$s_{mm}$ is negative, in order to have an entropy maximum, this
indirectly proves that a canonically stable solution is also
microcanonically stable. The converse is not true: one may well
have an entropy maximum, $s_{mm}<0$, which is a free energy saddle
point, with $s_{ee}>0$. This implies that the specific
heat~(\ref{cv})
can be negative \index{negative!specific heat}.\\
The above results are actually quite general, provided the
canonical and microcanonical solutions are expressed through
variational problems of the type (\ref{canoproblem}) and
(\ref{microproblem}).
The extrema, and thus the caloric curves \index{caloric curve} $T(e)$,
are the same in the
two ensembles, but the stability of the different branches is
different (see also the important paper by Katz~\cite{Katz} and
the contribution by Chavanis to this book~\cite{Chavanis} for
a discussion of this point in connection to self-gravitating
systems).
Another example of this behavior is discussed for the HMF model
in this book~\cite{HMF}.

\section{Maxwell constructions}
\label{Section4}

In this Section we briefly comment on the Maxwell construction
\index{Maxwell!construction} leading to temperature jumps in the
microcanonical ensemble. This is quite a new feature, a sort of
analogue of the latent heat phenomenon in the canonical ensemble.

\begin{figure}
\centering

\psfrag{B}{$\beta$}
\psfrag{Bc}{$\beta_c$}
\psfrag{e1}{$e_1$}
\psfrag{e2}{$e_2$}
\psfrag{e3}{$e_3$}
\psfrag{A1}{$A_1$}
\psfrag{A2}{$A_2$}
\psfrag{e}{$e$}
\psfrag{a}{a)}

\psfrag{ec}{$e_c$}
\psfrag{B1}{$\beta_1$}
\psfrag{B2}{$\beta_2$}
\psfrag{B3}{$\beta_3$}
\psfrag{M1}{$A_1$}
\psfrag{M2}{$A_2$}
\psfrag{b}{b)}
\includegraphics[scale=.5]{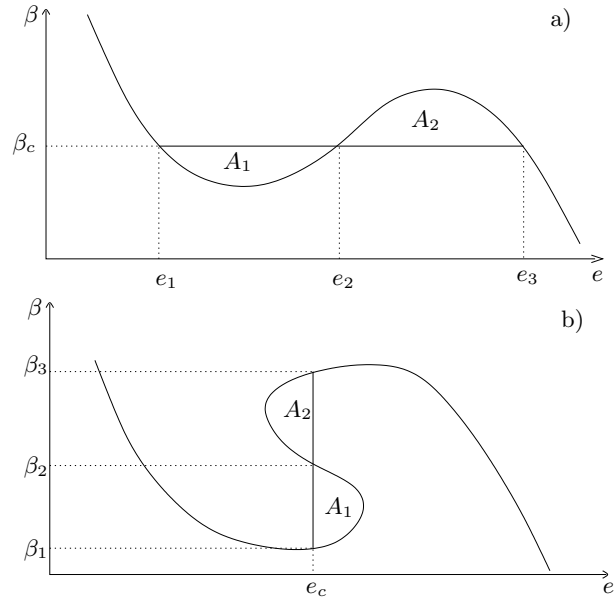}

\caption{Canonical (a) and microcanonical (b) Maxwell
constructions.} \label{maxwell}
\end{figure}

Let us begin with the ordinary Maxwell construction in the
canonical ensemble. We consider the curve $\beta(e)$ as in
Fig.~\ref{maxwell}a. It is the situation of Fig.~\ref{tvse}c, if
for simplicity we disregard the discontinuity in the temperature
derivative, which is irrelevant for the present reasoning. The
equal area Maxwell condition $A_1=A_2$ simply reads
\begin{equation}
\int_{e_1}^{e_3}[\beta(e)-\beta_c]\:de=0~,
\end{equation}
where $\beta_c$ is the first order transition inverse temperature.
Since $\beta=ds/de$, this implies
\begin{equation}
s(e_3)-s(e_1)-\beta_c(e_3-e_1)=0.
\end{equation}
Using now the definition of free energy, $f=e-s/\beta$, we obtain
the usual equal free energy condition at a first order phase
transition,
\begin{equation}
f(e_1)=f(e_3)~.
\end{equation}

Near a discontinuous microcanonical transition, the $\beta(e)$
curve has the typical shape given in Fig.~\ref{maxwell}b. This is
in fact the case of Fig.~\ref{tvse}, if one continues the low
energy branch of $\beta(e)$ above the microcanonical transition
energy, drawing the metastable and the unstable lines as well. We
invert this relation to obtain $e(\beta)$ in the vicinity of
$e_c$. Then, the equal areas condition reads
\begin{equation}
\int_{\beta_1}^{\beta_3}[e(\beta)-e_c]\:d\beta=0~.
\label{areamicro}
\end{equation}
using $e=d(\beta f)/d\beta$, we get
\begin{equation}
\beta_3 f(\beta_3)-\beta_1f(\beta_1)-e_c(\beta_3-\beta_1)=0~,
\end{equation}
which implies
\begin{equation}
s(\beta_3)=s(\beta_1)~.
\end{equation}
This is the equal entropy condition at a microcanonical
discontinuous transition. Eq.~(\ref{areamicro}) is valid only for
the simple configuration of Fig.~\ref{maxwell}b; for more complex
curves one has to evaluate the areas more carefully (see also the
discussion by Chavanis~\cite{Chavanis}).

\section{The microcanonical solution by large deviation techniques}
\label{Section5} \index{microcanonical!ensemble}

In Section~\ref{Section3} we derived an expression for the
logarithm of the density of states at fixed energy and
magnetization, i.e. the entropy $s(e,m)$, using combinatorial
techniques. We now present an alternative derivation of the
entropy using Cram{\'e}r's theorem~\cite{dembo}. This theorem,
derived in the context of large deviation theory \index{large
deviation theory}, states the following: consider a set of $N$
independent and identically distributed random variables in $R^d$,
$(X_i)_{i=1\ldots N}$.  One would like to know the probability
distribution of the average $M_N=1/N \sum X_i$. For this purpose,
one defines, for $\lambda \in R^d$,
\begin{equation}
\psi(\lambda)=\ln{\langle e^{\lambda \cdot X}\rangle},
\end{equation}
where the average $\langle\rangle$ is taken with respect to the
common Probability distribution Function (PdF) of all $X_i$. Now
let $s(x)/k_B$, with $x\in R^d$, be the inverse Legendre transform
of $\psi$
\begin{equation}
\label{cramerentropy}
\frac{s(x)}{k_B}=-\sup_{\lambda \in R^d}
\left(\lambda \cdot x - \psi(\lambda)\right)~.
\end{equation}
Then, under quite general conditions, Cram{\'e}r's theorem states
that the PdF of the average is
\begin{equation}
P(M_N=x) \sim e^{N\frac{s(x)}{k_B}}~.
\end{equation}
This yields an expression for the logarithm of the density of
states $s(x)/k_B$. The great advantage of this method is, of course,
its generality, since it applies also when combinatorial tools are
not available.

Let us give a brief heuristic argument for this result in the
$d=1$ case, $\lambda,x \in R$. The probability of having $M_N=x$
is given by the volume in phase space compatible with $M_N=x$. Let
$d\mu$ be the common measure of all $X_i$. One gets,
\begin{eqnarray}
P(M_N=x)  & = & \int d\mu(X_1)\ldots d\mu(X_N) \delta(M_N-x)\\
          & = & \frac{1}{2\pi i}\int_{\Gamma} d\lambda e^{-N\lambda x}
      \int d\mu(X_1)
      \ldots d\mu(X_N) e^{\lambda(X_1+\ldots+X_N)}\\
  & = & \frac{1}{2\pi i} \int_{\Gamma} d\lambda e^{-N\lambda x} \left[\langle
  e^{\lambda X} \rangle \right]^N~,
\end{eqnarray}
where the Dirac $\delta$ function has been represented by a Laplace
integral over a path $\Gamma$ transverse to the real axis
in the complex $\lambda$ plane. Evaluating the last integral using
the saddle point method one is lead to look for the critical
points of $\lambda x-\psi(\lambda)$, with $\psi(\lambda)=\ln
[\langle e^{\lambda X} \rangle]$, which justifies Eq.~(\ref{cramerentropy}).\\

As an example, let us apply this method to the BEG model\footnote{
An application to the Ising model and to other systems can be
found in~\cite{ellisrevue}}. In this case the $X_i$ variables are
bidimensional ($d=2$), $X_i=(S_i^2,S_i)$, so that
\begin{eqnarray}
\psi(\lambda,\rho)&=&\ln{\langle e^{\lambda S_i^2 + \rho S_i}\rangle} \nonumber \\
                 &=&\ln(1+2e^{\lambda}\cosh{\rho})-\ln 3
\end{eqnarray}
To calculate $s(q,m)$ we now have to solve equations $\partial
\psi/\partial\lambda=q$ and  $\partial \psi/\partial \rho=m$ for
$\lambda$ and  $\rho$. One gets
\begin{eqnarray}
\lambda=\ln \frac{q\sqrt{1-r^2}}{2(1-q)} \\
\rho= \ln \sqrt{\frac{1+r}{1-r}}
\end{eqnarray}
where $r$ is the ratio $m/q$. Substituting in (\ref{cramerentropy}),
we obtain the expected result
\begin{equation}
\frac{s(q,m)}{k_B}=-\frac{q+m}{2}\ln \frac{q+m}{2}-\frac{q-m}{2}\ln \frac{q-m}{2}
-(1-q)\ln(1-q)-\ln 3~,
\end{equation}
where the $-\ln 3$ constant is a normalization factor related to the fact
that the total number of configurations is $3^N$.

\section{Slowly decaying interactions}
\label{Section6}

So far, we have restricted our study to infinite-range models.
Space is in this case irrelevant, which simplifies very much the
study. However, the physically interesting interactions (gravity
for instance) are not infinite range, and one may wonder whether
the properties of the previous Sections still hold in a more
general case. To answer this question we  discuss in this Section
spin systems on $d$-dimensional lattices with pairwise interaction
potential between two sites $i$ and $j$ decaying with the distance
$r_{ij}$ like $1/r_{ij}^{\alpha}$ ($0 \leq \alpha < d$), so that
the interaction is long-range. An Ising model with short range
algebraically decaying interactions ($\alpha > d$) has been
studied in the past by Dyson~\cite{dyson}. More recently, versions
of the Hamiltonian Mean Field model~\cite{HMF}, with slowly
decaying long-range interactions among rotors ($\alpha < d$) have
been considered by several authors~\cite{anteneodo,campamm} and
the problem has been discussed in a more general framework in
Ref.~\cite{luijten}.

Our main result is that all peculiar features of the
microcanonical ensemble of infinite range models extend to these
slowly decaying interactions. We proceed by considering a
continuum, or coarse grained version of magnetic systems. We then
explain that in fact these coarse grained models can be derived
from the microscopic models, like the Ising or the BEG models
considered in this paper, using the large deviations tools introduced
in Section~5 (see \cite{FreddyJulien} for a more thorough discussion
of this point).

We begin by analyzing the phase diagram of the continuum version
of the Ising model with algebraically decaying interactions
$(0<\alpha < d)$. We show that as in the mean field model ($\alpha
=0$) this model does not exhibit a phase transition in the
microcanonical ensemble. We then consider the continuum version of
the BEG model.

The Ising Hamiltonian functional with algebraically decaying
interactions in $d=1$ dimensions on the segment $[0,1]$ with
periodic or free boundary conditions (to be discussed below) takes
the form
\begin{equation}
h_{Ising} = -\frac{{\bar J}}{2} \int_0^1 dx \int_0^1 dy \frac{m(x) m(y)}
{r(x,y)^{\alpha}},
\label{approxIsing}
\end{equation}
where $m(x)$ is the local magnetization, $r(x,y)$ is the distance
between $x$ and $y$ on the segment (the shortest distance for
periodic boundary conditions), and ${\bar J}>0$ is a ferromagnetic
coupling constant . The entropy density corresponding to a
magnetization profile $m(x)$ of an Ising variable may be written
as
\begin{equation}
\label{Entropy1Ising}
s_{Ising}=-k_B \int_0^1 dx[\frac{(1+m)}{2} \ln \frac{(1+m)}{2}
+\frac{(1-m)}{2} \ln \frac{(1-m)}{2}]~,
\end{equation}

Once the energy and entropy functionals are given, the
microcanonical (canonical) solution is obtained by a maximization
(minimization) of the entropy (free-energy) functional under the
constraint of constant energy (temperature). We begin below with
the discussion of these two variational problems for the Ising model
\begin{equation}
s_{Ising}(e) = \max_{m(x)} \left(s_{Ising}[(m(x)] \:
\left| \quad h_{Ising}=e
\right. \right),
\label{microsolution}
\end{equation}
in the microcanonical ensemble \index{microcanonical!ensemble},
and
\begin{equation}
f_{Ising}(\beta) = \min_{m(x)} \left( h_{Ising}[m(x)]
-\frac{1}{\beta} s_{Ising}[m(x)] \right) \label{canosolution}
\end{equation}
in the canonical ensemble \index{canonical!ensemble}.

So far, although we have stated the problem for the Ising model,
we could have followed the same path, whatever the lattice model and the
boundary conditions are. We can reach a general conclusion at
this point by writing down the extremality conditions of the two
variational problems (\ref{microsolution}) and
(\ref{canosolution}). The first order reads
\begin{equation}
\frac{\delta s_{Ising}}{\delta m(x)} =  \beta
\frac{\delta h_{Ising}}{\delta m(x)}~,
\label{microorder1}
\end{equation}
in the microcanonical ensemble, where $\beta$ is a Lagrange multiplier.
In the canonical ensemble
\begin{equation}
\frac{\delta h_{Ising}}{\delta m(x)} - \frac{1}{\beta}
\frac{\delta s_{Ising}}{\delta m(x)}=0~.
\label{canoorder1}
\end{equation}
Hence, the first order extremality conditions lead to the same
equations for the two ensembles. However, this \emph{does not}
imply ensemble equivalence, since the stability of the solutions
may differ in the two ensembles, as has been noted in
Section~\ref{Section4}.

Before proceeding, let us briefly recall the results of
Ref.~\cite{anteneodo}. These authors, discussing the Hamiltonian
Mean Field model with slowly decaying
interactions~\cite{HMF,Tsallis},  remarked that, choosing
appropriately the renormalization factor in the Hamiltonian, the
thermodynamic behavior of the system turned out to be independent
of $\alpha$; i.e. the system behaves at $0<\alpha<1$ (slowly
decaying case) exactly as if $\alpha=0$ (mean-field case). An
explanation of this result was provided by Campa et
al.~\cite{campamm} through a canonical analysis of long-range
interacting systems on lattices with periodic boundary conditions, and
independently by Vollmayr-Lee and Luijten~\cite{luijten}
in a more general context.
Unfortunately, neither of the two latter methods provide the microcanonical
solution. The method we present here is, instead, fully general
and allows to discuss in detail how to extend the results for
$\alpha=0$ to $\alpha>0$.

Let us solve now the variational problem of
Eqs.(\ref{microorder1}) and (\ref{canoorder1}) for the Ising
model~\cite{Barre}. The magnetization profile that extremizes
entropy and free energy satisfies the following self-consistency
condition
\begin{equation}
m(x)=\tanh \left(\beta \bar{J}\int_0^1\frac{m(y)}{r(x,y)^{\alpha}}\:dy\right)
\label{eqising}
\end{equation}
In the canonical ensemble, $\beta$ is fixed, whereas in the
microcanonical ensemble it has to be tuned in order to get the
required energy. For periodic boundary conditions, the integral
on the r.h.s. of Eq.~(\ref{eqising})
does not depend on $x$, and Eq.(\ref{eqising}) always has a
homogeneous solution with $m(x)$ independent of $x$.
Eq.~(\ref{eqising}) then reduces to the usual consistency equation
of the mean-field Ising model, and the system behaves exactly as
in the $\alpha=0$ case, showing a phase-transition in the canonical
ensemble (no transition is present in the microcanonical
ensemble).

\begin{figure}
\centering
\psfrag{position on the lattice}[Bl][Bl][0.5][0]{Position on the lattice}
\psfrag{magnetization}[Bl][Bl][0.5][0]{Magnetization}
\psfrag{mag profile for different alpha value}{}
\psfrag{mag profile for periodic boundary conditions}{}
\psfrag{0}[][][.5][0]{$0$}
\psfrag{1}[][][.5][0]{$1$}
\psfrag{0.5}[][][.5][0]{$0.5$}
\psfrag{0.1}{}
\psfrag{0.2}{}
\psfrag{0.3}{}
\psfrag{0.4}{}
\psfrag{0.6}{}
\psfrag{0.7}{}
\psfrag{0.8}{}
\psfrag{0.9}{}
\includegraphics[width=.5\textwidth]{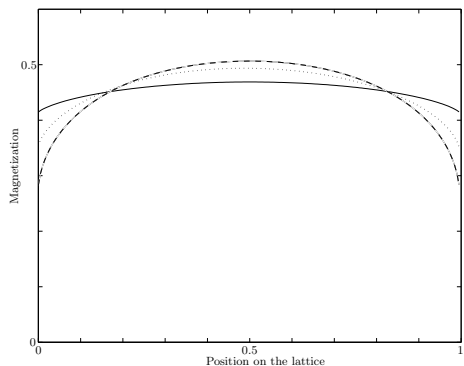}

\caption{Magnetization profile for the Ising model with slowly decaying
interactions in the microcanonical ensemble. The energy density is fixed
at $e=-0.1$
while $\alpha$ varies: $\alpha=0.2$ (full line), $\alpha=0.5$ (dotted
line), $\alpha=0.8$ (dashed line). The boundary conditions are free.}
\label{profile}
\end{figure}

On the contrary, for free boundary conditions, there is no
non-zero constant solution, and the system develops a non trivial
magnetization profile that can be only numerically computed from
Eq.(\ref{eqising}). Moreover, in this case a phase transition is
present in the canonical ensemble, while the microcanonical
ensemble is always in a magnetically ordered phase. The
magnetization profile in the microcanonical ensemble for free
boundary conditions is shown in Fig.~\ref{profile} for three
different values of $\alpha$.

Let us now discuss the behavior of the BEG model for slowly decaying
interactions. The energy functional is
\begin{equation}
h_{BEG} = \Delta \int_0^1 dx q(x) - \frac{{\bar J}}{2}
\int_0^1 dx \int_0^1 dy \frac{m(x) m(y)}{r(x,y)^{\alpha}},
\label{approxbeg}
\end{equation}
where all is defined as before and, moreover, $q(x)$ is the
quadrupolar field in the continuum on the segment $[0,1]$
(obtained in an analogous way as for the magnetization $m(x)$).
The entropy functional of the BEG model is obtained from
expression~(\ref{Entropy1}) by replacing $m,q$ by $m(x),q(x)$.

The BEG model is much richer, as has already been demonstrated for
$\alpha=0$. The analysis of the microcanonical phase diagram for
periodic boundary conditions follows the same path as that
presented above for the Ising model. It is straightforward to show
that the constant profiles $m(x)=m^{\ast}$, $q(x)=q^{\ast}$,
with $m^{\ast}$ and $q^{\ast}$ given by the solution of the
$\alpha=0$ model, are solutions of the Eqs.~(\ref{microorder1})
and~(\ref{canoorder1}) (where we replace the Ising subscript by
the BEG one) also for $\alpha>0$. Moreover, detailed
calculations~\cite{BMR2} show that these solutions are local
maxima of the entropy and local minima of the free energy. Thus
one concludes that for periodic systems the entropy and free
energy are independent of $\alpha$ and the conclusions reached for
$\alpha=0$ are valid for $0< \alpha <d$. In particular, ensemble
inequivalence \index{ensemble inequivalence} can be extended to
the case of long-range algebraically decaying interactions.

Let us briefly sketch these calculations. To verify that uniform
profiles are not destabilized by modulations, it is convenient to
develop $q(x)$ and $m(x)$ in Fourier series:
$q(x)=\sum_k\tilde{q}_k u_k(x)$, $m(x)=\sum_k\tilde{m}_k u_k(x)$,
where $k$ ranges from $-\infty$ to $+\infty$ and $u_k(x)=\cos kx$
if $k \geq 0$, $u_k(x)=\sin kx$ if $k < 0$ .

The Fourier representation, due to the periodicity, diagonalizes
the Hamiltonian. The energy may thus be written as
\begin{equation}
\epsilon=\tilde{q}_0 - {\bar K} \sum_{k=-\infty}^{+\infty}\lambda_k \tilde{m}_k^2~.
\end{equation}
Here $\lambda_k$ are the energy eigenvalues and ${\bar
K}=2^{\alpha}\bar{J}/2\Delta(1-\alpha)$ is chosen such that the
largest eigenvalue is unity $\lambda_0=1$ (this amounts to choose
the prefactor $\tilde{N}$ in the Kac prescription, see
below)\index{Kac prescription}.

The free energy functional is
\begin{equation}
f[q,m]= \epsilon-1/\beta \int s[q(x),m(x)]\:dx~,
\end{equation}
so that we get the first order conditions
\begin{eqnarray}
&&\delta_{k0}\beta-\int s_q[q,m]u_k(x)\:dx = 0 \\
&&-2\beta {\bar K}\lambda_k\tilde{m}_k-\int s_m[q,m]u_k(x)\:dx = 0
\end{eqnarray}
We can easily check that the uniform profile $q=\tilde{q}_0$,
$m=\tilde{m}_0$ (with $\tilde{q}_0$ and $\tilde{m}_0$ given by
the $\alpha=0$ case) is indeed a solution of these
conditions. To prove that the uniform profile is a free energy
minimum, we evaluate now the Hessian matrix at this point
\begin{eqnarray}
\frac{\partial f}{\partial \tilde{q}_k\partial \tilde{q}_l} & = &
-\frac{s_{qq}}{\beta}\delta_{kl}\\
\frac{\partial f}{\partial \tilde{q}_k\partial \tilde{m}_l} & = &
-\frac{s_{qm}}{\beta}\delta_{kl}\\
\frac{\partial f}{\partial \tilde{m}_k\partial \tilde{m}_l} & = &
-\frac{s_{mm}}{\beta}\delta_{kl}-2{\bar K}\lambda_k\delta_{kl}~.
\end{eqnarray}
This infinite Hessian matrix turns out to be positive
definite. To prove this, it is sufficient to verify that each matrix
$H_k$
\begin{equation}
H_k= -\left( \begin{array}{ll} s_{qq} & s_{qm} \\
                              s_{qm} & s_{mm}+2{\bar K}\beta\lambda_k
\end{array} \right)
\end{equation}
is positive definite, when taken at the solution point. This follows easily
from the fact that $H_0$
\begin{equation}
H_0= -\left( \begin{array}{ll} s_{qq} & s_{qm} \\
                              s_{qm} & s_{mm}+2{\bar K}\beta
\end{array} \right)
\end{equation}
is positive definite (because the point considered is the solution of the
$\alpha=0$ case), and recalling that $\lambda_k\leq1$ for whatever $k$.

We have thus proved that the local stability of the uniform
profiles at $\alpha=0$ are not modified in the canonical ensemble
when $\alpha > 0$. All canonical $T(e)$ curves shown in
Fig.~\ref{tvse} thus extend to the $\alpha>0$ case. We now make
use of the general discussion at the end of Section \ref{Section3}.
We recall that results there derived state, in particular, that a
stable canonical solution is always a stable microcanonical one,
and that if one follows a $T(e)$ curve, like in Fig.~\ref{tvse}, a
change of stability in the microcanonical ensemble may occur only
where this curve has a vertical tangent (see~\cite{Katz} and the
contribution by Chavanis~\cite{Chavanis} to this book). We
thus conclude that the microcanonical part of Fig.~\ref{tvse} does
not change either. This reasoning is however not rigorous, since
we have considered only local stability; it may happen that when
$\alpha > 0$, a heterogeneous solution appears and it could yield
the true free energy minimum or entropy maximum. Some numerical
investigations we have carried out did not detect any such
heterogeneous solution.

We now turn to a different issue, namely the correspondence
between the microscopic models and the continuum ones. In the
following we show that the Ising energy functional
(\ref{approxIsing}) can be derived by coarse graining from the
microscopic Ising Hamiltonian with slowly decaying interactions
\begin{equation}
H_{Ising}= -\frac{J}{2\tilde{N}}\sum_{i,j}
\frac{\sigma_i\sigma_j}{r_{ij}^{\alpha}}~,
\label{alphaising}
\end{equation}
where $r_{ij}$ is the distance on a 1D lattice between spins at
sites $i$ and $j$. The interaction is {\it non integrable}
for $\alpha \leq 1$. The normalization
$\tilde{N}=2^{\alpha}N^{1-\alpha}/(1-\alpha)$ ensures that the energy is
extensive \index{extensivity} \index{Kac prescription}
and the prefactor provides a convenient normalization for the
eigenvalues of the Hamiltonian in the case of periodic boundary conditions.
To make the correspondence between continuous and microscopic
models explicit, one has to set $\bar{J}=(1-\alpha)J/2^{\alpha}$
in Eq.~\ref{approxIsing} .

Analogously, the BEG microscopic model we refer to is
\begin {equation}
H_{BEG}= \Delta \sum_{i=1}^{N} S_i^2
-\frac{J}{2\tilde{N}}\sum_{i,j} \frac{S_iS_j}{r_{ij}^{\alpha}}~.
\label{alphabeg}
\end{equation}

The coarse-graining procedure we present, closely follows
Ref.~\cite{ellis}. We will keep here a heuristic level of
description, although all the following calculations can be made
rigorous using the language and techniques of large deviation
theory~\cite{dembo}. A more detailed account of this derivation is
given in Ref.~\cite{FreddyJulien}.

The first step involves a coarse-graining procedure. We divide the
lattice in $N/n$ boxes, each one containing $n$ sites. We then
describe macroscopically the system in the Ising case by an average
magnetization in each box, or, in the BEG case, by the average $m$
and $q$. In the limit $N \to \infty$, $n \to \infty$
with $N/n \to \infty$, the system is then described by
continuous functions $m(x),q(x)$,
with $x\in[0,1]$ if the system length is normalized. We have to
show in the following that all the information lost in the coarse
graining procedure is unessential.

In the mean-field case ($\alpha=0$), the Hamiltonian had a natural
and exact expression as a function of the macroscopic parameters
$m$ (Ising) or $m$ and $q$ (BEG). The long-range Ising case has
already been discussed~\cite{Barre} and it has been shown that the
intra-box couplings can be neglected in the thermodynamic limit,
leading to the expression for the Hamiltonian functional
(\ref{approxIsing}) and for the entropy (\ref{Entropy1Ising}) as a
function of the macroscopic parameter  $m(x)$. The generalization
to the BEG model of the method applied in Ref.~\cite{Barre} is
straightforward and allows one to obtain the Hamiltonian
functional (\ref{approxbeg}). The estimation is
uniform over all the microscopic configurations~\cite{FreddyJulien},
and $r(x,y)$ is
the distance between the two boxes located in $x$ and $y$. On the
contrary, if $\alpha>d$, there is no way to approximate the
Hamiltonian in the continuum limit described above as a functional
of $m(x)$ and $q(x)$.

We show now how to construct the entropy functional for the Ising
model. Its extension to the BEG model is straightforward. We have
to estimate the probability to obtain a certain given macroscopic
configuration $m(x)$, assuming the equiprobability of microscopic
configurations. For a finite number of boxes, $N/n$, we can evaluate
the probability to get a given average magnetization in each box
\begin{eqnarray}
P(m_1,m_2,\ldots,m_n) & = & P(m_1)P(m_2)\cdots P(m_n) \nonumber\\
& \simeq & e^{ns_1(m_1)/k_B}\cdots e^{ns_n(m_n)/k_B}
\end{eqnarray}
where $s_i$ is the entropy associated to the $i$-th box. Letting $n$ and
$N/n$ go to infinity, we obtain
\begin{equation}
P[m(x)] \simeq  e^{\frac{N}{k_B}\int_0^1 s(m(x))\: dx}~,
\label{largedevising}
\end{equation}
which defines the entropy functional $s_{Ising}$ of the Ising
model. Let us remark again that large deviation techniques~\cite{dembo}
give a precise and rigorous meaning to these calculations,
and allow one to extend this type of derivation to all sort of
lattice models. The field $q(x)$ of the BEG model can be treated
in a similar way as $m(x)$ and one gets the BEG entropy
functional similarly.

Chavanis~\cite{Chavanis} and, Cohen and Ispolatov~\cite{Cohen} analyze in this
book more realistic off-lattice systems, self-gravitating or
interacting through a $1/r^{\alpha}$ potential
($\alpha<3$)\index{astrophysics} \index{gravitation}. They study the
thermodynamic properties of these models
through a mean-field approximation, and the phenomenology is pretty
much the same as the one decribed here. Unfortunately, to see the
analogies, one must cope with discrepancies in the ``vocabulary".
For instance, the {\it microcanonical first order} transition described
here is called \emph{gravitational first order} transition, whereas
the {\it canonical first order} one is
called \emph{normal first order} transition. Within the
conventions of this paper one can see that, varying a control parameter (the
interaction $K$ for the BEG model, and a parameter controlling
the short range cut-off in gravitational models) the system crosses
over from a microcanonical first order transition to a microcanonical
second order one (or no transition at all in the gravitational systems)
associated with a canonical first order, and then
recovers full ensemble equivalence, crossing the critical point.

All these off-lattices studies raise the question of the validity
of the mean-field approximation, which is intimately related to
the scaling with $N$ of the thermodynamic variables and
potentials. In our case, the mean field treatment is fully valid
in the thermodynamic limit, but the problem was avoided using the
physically unjustified Kac prescription~\cite{Kac} in which the
coupling constants are rescaled by $\tilde{N}$.

\section{Conclusions}
\label{Section7}

In this paper a detailed comparison is made between the canonical
and the microcanonical phase diagrams of the spin-1 BEG model with
mean field long range interactions. Since systems with long range
interactions, and particularly mean field models, are non-additive
the two ensembles need not be equivalent. The BEG model, for which
both phase diagrams can be evaluated analytically, provides a
convenient and interesting ground for studying the distinctions
between the two ensembles. Although the model is rather simple, it
exhibits a non-trivial phase diagram, which in the canonical
ensemble consists of both a first order and a second order
transition lines separated by a tricritical point. It is shown that the
microcanonical phase diagram, while it yields the same
thermodynamic behavior in the region where the canonical
transition is second order, it differs considerably from the
canonical one in the region where the transition is first order.
In particular it exhibits a region with a negative specific heat,
a tricritical point whose location differs from that of the
canonical ensemble, and a first order line at which the two
coexisting phases have different temperatures. The mechanisms
which lead to these features are rather general, and thus they are
expected to take place in other models with long range
interactions within the microcanonical ensemble (see
also~\cite{Leyvraz,Ispolatov}).

A straightforward generalization of these results to models with
algebraically decaying interactions are considered. It is shown
that the mean field phase diagram remains valid even for these
interactions, as long as they are long range.

It would be of great interest to extend this study in the future
and to analyze and compare canonical and microcanonical phase
diagrams in more complex cases, where higher order multicritical
points are present. Such studies may eventually lead to general
rules which should be obeyed by microcanonical phase diagrams of
systems with long range interactions, in analogy to the rules
existing for systems with short range interactions, like the Gibbs
phase rules, the Landau and Lifshitz symmetry rules for continuous
transitions and others.

\section*{Acknowledgements}
We would like to warmly thank our collaborators Freddy Bouchet and
Fran{\c c}ois Leyvraz for fruitful interactions.  We thank Thierry Dauxois
and Dieter Gross for very useful discussions. This work has been
partially supported by the EU contract No. HPRN-CT-1999-00163 (LOCNET
network) and the R{\'e}gion Rh{\^o}ne-Alpes for the fellowship
N$^\circ$ 01-009261-01.  This work is also part of the contract
COFIN00 on {\it Chaos and localization in classical and quantum mechanics}.

%INDEX%%%%%%%%%%%%%%%%%%%%%%%%%%%%%%%%%%%%%%%%%%%%%%%%%%%%%%%%%%%%%%%
% Please check with the editor of your book whether he plans to
% include a "mutual" subject index - if so, please code your entries
% in the standard syntax. For your own purposes you may print your
% "personal" index by using the following commands:
%
%\clearpage
%\addcontentsline{toc}{section}{Index}
%\flushbottom
%\printindex
%%%%%%%%%%%%%%%%%%%%%%%%%%%%%%%%%%%%%%%%%%%%%%%%%%%%%%%%%%%%%%%%%%%%%


\begin{thebibliography}{8.}
\addcontentsline{toc}{section}{References}

\bibitem{Kac} M.~Kac, G.E.~Uhlenbeck and P.C.~Hemmer, J. Math. Phys. {\bf 4},
216 (1963).

\bibitem{Antonov} V.A.~Antonov Vest. Leningrad Univ. {\bf 7}, 135
(1962); Translation in IAU Symposium {\bf 113}, 525 (1995).

\bibitem{Lynden} D.~Lynden-Bell and R. Wood, Mon. Not. R. Astron. Soc. {\bf 138},
495 (1968); D.~Lynden-Bell, Physica A {\bf 263}, 293 (1999).

\bibitem{Thirring} W.~Thirring, {\it Z. Phys.} {\bf 235}, 339 (1970);
P.~Hertel and W.~Thirring, {\it Ann. of Phys.} {\bf 63}, 520 (1971).

\bibitem{Chavanis} P.H. Chavanis, \emph{Statistical mechanics of
    two-dimensional vortices and three-dimensional stellar systems},
  in ``Dynamics and Thermodynamics of Systems with Long Range
  Interactions'', T.  Dauxois, S. Ruffo, E. Arimondo, M. Wilkens Eds.,
  Lecture Notes in Physics Vol. 602, Springer (2002), (in this volume)
  
\bibitem{Padhmanaban} T. Padhmanaban, {\it Statistical mechanics of
    gravitating systems in static and cosmological backgrounds}, in
  ``Dynamics and Thermodynamics of Systems with Long Range
  Interactions'', T.  Dauxois, S. Ruffo, E. Arimondo, M. Wilkens Eds.,
  Lecture Notes in Physics Vol. 602, Springer (2002), (in this volume)
  
\bibitem{Gross} D.H.E.~Gross, {\it Microcanonical thermodynamics:
    Phase transitions in ``small" systems}, (World Scientific,
  Singapore, 2000) and {\it Thermo-Statistics or Topology of the
    Microcanonical Entropy Surface}, in ``Dynamics and Thermodynamics
  of Systems with Long Range Interactions'', T.  Dauxois, S. Ruffo, E.
  Arimondo, M. Wilkens Eds., Lecture Notes in Physics Vol. 602,
  Springer (2002), (in this volume)

\bibitem{RMBell} R.M.~Lynden-Bell, in {\it Gravitational dynamics},
O.~Lahav, E.~Terlevich and R.J.~Terlevich (eds.), Cambridge Univ.
Press (1996); R.M.~Lynden-Bell, Mol. Phys. {\bf 86}, 1353 (1995).

\bibitem{Chomaz} P. Chomaz and F. Gulminelli {\it Phase transitions
in finite systems},  in
  ``Dynamics and Thermodynamics of Systems with Long Range
  Interactions'', T.  Dauxois, S. Ruffo, E. Arimondo, M. Wilkens Eds.,
  Lecture Notes in Physics Vol. 602, Springer (2002), (in this volume)


\bibitem{Blume} M.~Blume, V.J.~Emery and R.B.~Griffiths, Phys. Rev. A
{\bf 4}, 1071 (1971).

\bibitem{BMR} J. Barr{\'e}, D. Mukamel and S. Ruffo, Phys. Rev. Lett.
{\bf 87}, 030601 (2001).

\bibitem{Katz} J.~Katz, Mon.~Not.~R.~Astr.~Soc. \textbf{183}, 765
(1978).

\bibitem{dembo} A.~Dembo, O.~Zeitouni, \emph{Large Deviations Techniques and
Applications}, (Springer, Berlin, 1998).

\bibitem{ellisrevue} R.~S.~Ellis, Physica D \textbf{133}, 106, (1999).

\bibitem{dyson} F.J. Dyson, Comm. Math. Phys., {\bf 12}, 91 (1969).

\bibitem{HMF} T. Dauxois, V. Latora, A. Rapisarda, S. Ruffo and A. Torcini,
{\it The Hamiltonian Mean Field Model: from Dynamics to Statistical Mechanics
and back},  in
  ``Dynamics and Thermodynamics of Systems with Long Range
  Interactions'', T.  Dauxois, S. Ruffo, E. Arimondo, M. Wilkens Eds.,
  Lecture Notes in Physics Vol. 602, Springer (2002), (in this volume)

\bibitem{anteneodo} C.~Anteneodo and C.~Tsallis, Phys. Rev. Lett. {\bf 80},
5313 (1998); F.~Tamarit and C.~Anteneodo, Phys. Rev. Lett. {\bf 84}, 208
(2000);

\bibitem{campamm} A. Campa, A.~Giansanti and D. Moroni, Phys. Rev. E {\bf 62},
303 (2000) and Chaos Solitons and Fractals, {\bf 13}, 407 (2002).

\bibitem{luijten} B.P. Vollmayr-Lee and E. Luijten, Phys. Rev. E {\bf 63},
031108 (2001) and Phys. Rev. Lett. {\bf 85}, 470 (2000).

\bibitem{ellis} R.~S.~Ellis, K.~Haven, B.~Turkington, J. Stat. Phys.
{\bf 101}, 999 (2000).

\bibitem{Barre} J. Barr{\'e}, Physica A, {\bf 305}, 172 (2002).
  
\bibitem{Tsallis} C. Tsallis, A. Rapisarda, V. Latora and F. Baldovin,
  {\it Nonextensivity: from low-dimensional maps to Hamiltonian
    systems}, in ``Dynamics and Thermodynamics of Systems with Long
  Range Interactions'', T.  Dauxois, S. Ruffo, E. Arimondo, M. Wilkens
  Eds., Lecture Notes in Physics Vol. 602, Springer (2002), (in this
  volume)

\bibitem{FreddyJulien} J. Barr{\'e} and F. Bouchet,
``Mean-Field justified by large deviations results in long-range
interacting systems''. Proceedings of the Confererence "Dynamics and
thermodynamics of systems with long range interactions", Les Houches,
France, February 18-22 2002, Eds. T. Dauxois, E. Arimondo, S. Ruffo, M. Wilkens,
published on http://www.ens-lyon.fr/$\sim$tdauxois/procs02/

\bibitem{BMR2} J. Barr{\'e}, D. Mukamel and S. Ruffo, to be published.
  
\bibitem{Cohen} E.G.D. Cohen, I. Ispolatov, {\it Phase transitions in
    systems with $1/r^\alpha$ attractive interactions}, in ``Dynamics and
  Thermodynamics of Systems with Long Range Interactions'', T.
  Dauxois, S. Ruffo, E.  Arimondo, M. Wilkens Eds., Lecture Notes in
  Physics Vol. 602, Springer (2002), (in this volume)


\bibitem{Leyvraz}  F. Leyvraz and S. Ruffo, J. Phys. A, {\bf 35}, 285 (2002)
and Physica A, {\bf 305}, 58 (2002).

\bibitem{Ispolatov} I. Ispolatov and E.G.D. Cohen, Physica A, {\bf 295},
475 (2001).

\end{thebibliography}
\end{document}